\documentclass[aps,prb,reprint,twocolumn]{revtex4-1}
\usepackage{amsmath}
\usepackage{siunitx}
\usepackage{braket}
\usepackage{mathtools}
\usepackage{amssymb}
\usepackage{graphicx}
\usepackage{bm}
\usepackage{here}
\usepackage{color}

\begin{document}

\title{Momentum selective optical absorption in triptycene molecular membrane}
\author{Masashi Akita$^1$, Yasumaru Fujii$^2$, Mina Maruyama$^2$, Susumu Okada$^2$ and Katsunori Wakabayashi$^{1,3,4}$}
\email{Corresponding author: waka@kwansei.ac.jp}
\affiliation{$^1$School of Science and Technology, Kwansei Gakuin University, Gakuen 2-1, Sanda 669-1337, Japan}
\affiliation{$^2$Graduate School of Pure and Applied Science, University of Tsukuba, Tsukuba 305-8571, Japan}
\affiliation{$^3$National Institute for Materials Science (NIMS), Namiki
1-1, Tsukuba 305-0044, Japan}
\affiliation{$^4$Center for Spintronics Research Network (CSRN), Osaka University, Toyonaka 560-8531, Japan}

\begin{abstract}
The optical properties of triptycene molecular membranes (TMMs) under the
linearly and circularly polarized light irradiation have been
theoretically studied. Since TMMs have the 
double-layered Kagome lattice structures for their $\pi$-electrons, i.e., tiling of
 trigonal and hexagonal-symmetric rings, the electronic band
 structures of TMMs have non-equivalent Dirac cones and perfect flat
 bands. By constructing the tight-binding model to describe the
 $\pi$-electronic states of TMMs, we have evaluated the optical absorption
 intensities and valley selective excitation of TMMs based on the Kubo formula.
It is found that absorption intensities crucially depend on both light
 polarization angle and the excitation position in momentum space, i.e., 
the momentum and valley selective optical excitation. The polarization dependence
 and optical selection rules are also clarified by using group theoretical analyses.
\end{abstract}
\maketitle
\section{Introduction}
Two-dimensional (2D) atomically thin materials have attracted
significant attention owing to their unique physical and chemical
properties, which are derived from the low dimensionality of electronic
systems.~\cite{Chhowalla,Butler}  
Graphene~\cite{Novoselov_2005} is one of the most prominent 2D
materials which shows high carrier mobilities,~\cite{dean2010bn} half-integer quantum Hall
effect,~\cite{novoselov2005two-dimensional,zhang2005} and superconductivity.~\cite{cao2018unconventional} 
Owing to the honeycomb network structure 
of sp$^{2}$ carbon atoms,~\cite{RevModPhys.81.109} the electronic states
of graphene near the Fermi energy are well described by using the
massless Dirac equation and possess conical energy dispersion at $K$
and $K^\prime$ points of hexagonal 1st Brillouin zone (BZ). These two
nonequivalent Dirac $K$ and $K^\prime$ points are mutually
related by time-reversal symmetry. The independence and degeneracy of the
valley degree of freedom owing to Dirac cones 
can be used to control the electronic states, 
i.e., valleytronics,~\cite{schaibley2016valleytronics,rycerz2007valley,xiao2007valley-contrasting,gunlycke2011graphene,PhysRevB.84.195408} which
is analogous to spintronics and advantageous for the ultra-low-power
consumption electronic devices.  
The idea of valleytronics is also applied to transition metal
dichalcogenide with honeycomb structure such as MoS$_2$
 and has been experimentally demonstrated that the electrons in
each valley can be selectively excited by circularly polarized light
irradiation.~\cite{mak2014the,Wang_2012,PhysRevLett.108.196802,Zeng,Kioseoglou,PhysRevB.86.081301,Wu_2013}

Besides the hexagonal lattice structures such as graphene, 
Kagome lattice, which has the trihexagonal tiling network,
is of interest, because it also possesses electronic energy band
structure with valley structures, together with a perfect flat energy band. 
Kagome lattice has been the intensive research subject of theoretical studies
because of its peculiar magnetic,~\cite{mielke1999exact,PhysRevLett.98.107204,Yan, Sachdev} 
transport~\cite{liu2010simulating} and topological
properties.~\cite{liu2009spin,guo2009topological,hatsugai2011zq,beugeling2012topological,ezawa2018higher-order,bolens2019topological} 
However, experimental fabrication of Kagome lattice especially
composed of sp$^2$ carbon atoms is considered to be difficult.  
Recently, the bottom-up synthesis of 2D materials has been extensively
investigated. Examples of this approach are surface metal-organic
frameworks (MOFs)~\cite{makiura2010surface} and 
surface covalent-organic frameworks (COFs).~\cite{colson2011oriented,spitler2011a}
It is also suggested that the electronic states of 
2D MOF consisting of $\pi$-conjugated nickel-bis-dithiolene\cite{kambe2013conjugated} can be modeled by
the tight-binding model of Kagome lattice with the spin-orbit
interactions as a candidate of topological insulators.~\cite{wang2013prediction} 

Here, we focus on the aromatic hydrocarbon
triptycene~\cite{bartlett1942triptycene1}
that is the three-dimensional (3D) propeller type structure as the building blocks
of polymerized triptycene molecular membranes (TMMs).~\cite{zhang2012triptycene-based,bhola2013a}
There are two types of cross-linked structure in TMMs according to the
bonding shape of each bridge, i.e., zigzag and armchair
types. The recent first-principles calculations based on
density functional theory (DFT) have shown that these TMMs are
thermodynamically stable and become semiconducting with 
multiple Kagome bands,~\cite{JPSJ.87.034704, Fujii2} 
i.e., several sets of graphene energy bands with a flat energy band.
Especially, these multiple Kagome bands provide a good platform of
selective excitation of electrons with specific momentum, i.e., at $K$
and $K^\prime$ points. However, the effect of light irradiation on the
optical transition has not been studied yet. 
\begin{figure*}[hbt]
  \begin{center}
    \includegraphics[width=\textwidth]{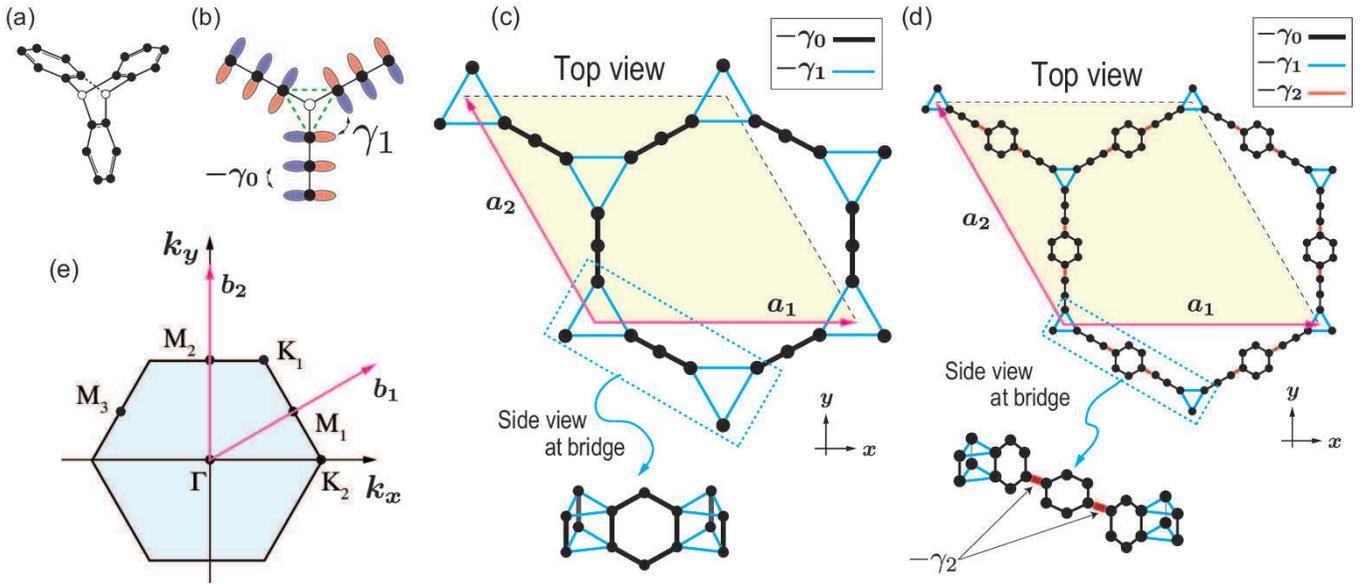}
  \end{center}
  \caption{
    (a) Schematic structure of triptycene molecule. The black and white
 carbons indicate sp$^2$ and sp$^3$ carbons, respectively. Each sp$^2$
 carbon gives one $\pi$-electron. (b) Top view of the triptycene
 molecule together with $\pi$-orbital at each sp$^2$ carbons.
$\pi$-orbitals construct the triangle network drawn with green dashed
 lines. Schematic structure of (c) zigzag TMM and (d) armchair TMM. 
The yellow shaded rhombus indicates the unit cell of TMMs. 
$\bm{a_1}=(a,0)$, $\bm{a_2}=(-a/2,\sqrt{3}a/2)$ are primitive vectors,
 where $a=8.92$\AA\ for zigzag TMM and $a=22.17$\AA\ 
for armchair TMM, respectively.
The zigzag and armchair TMMs have $18$ and $54$ $\pi$-electronic sites
 in their unit cells, respectively.
(e) 1st BZ of TMMs.
  }
  \label{fig:figure1}
\end{figure*}

In this paper, we theoretically study the optical properties of TMMs
under the linearly and circularly polarized light irradiation. 
To analyze the optical properties of TMMs, we construct the
tight-binding model that faithfully reproduces the energy band structure
obtained by DFT, and numerically evaluate the optical absorption
intensity and valley selective optical excitation using the Kubo formula.~\cite{kubo1957jpsj,mahan} 
It is found that absorption intensity crucially depends on both light
 polarization angle and the momentum of optically excited electrons.
It is also confirmed that the circularly polarized light irradiation can
selectively excite the electrons in either $K$ or $K^\prime$ point. 
Then, we analyze the optical selection rule of TMMs
using the group theory. From the analysis, we determine the 
selection rules of the absorption spectrum and polarization-dependent
transition at the high symmetric points in 1st BZ. 

This paper is organized as follows. In Sec. II, we illustrate the 
lattice structures of zigzag and armchair TMMs. Subsequently, 
we construct the tight-binding model
to calculate their electronic structures and describe the fundamental
theoretical framework to evaluate the optical properties of TMMs.
In Sec. III, we discuss the optical properties of zigzag TMM under
linearly and circularly polarized light irradiation.
The optical properties of armchair TMM is discussed in Sec. IV.
Section V summarizes our results. In addition, 
the details of the optical selection rules for 2D Kagome lattice are given in the Appendix~\ref{apeA}. 

\section{Tight-binding Model of TMM}
In this paper, we employ the tight-binding
model to describe the $\pi$-electronic states of TMMs and study their optical 
properties under the linearly and circularly polarized light
irradiation. 
The Hamiltonian for the $\pi$-orbitals in TMMs can be
given as 
\begin{equation*}
  \hat{H} = \sum_{\langle n,m\rangle}\gamma_{n,m} \ket{\psi_n}
   \bra{\psi_m},
\end{equation*}
where $\psi_m$ and $\psi_n$ indicate the $\pi$-orbitals at $m$ and $n$ sites in the unit
cell, respectively. $\gamma_{n,m}$ indicates the $\pi$-electron hopping integral between
$m$ and $n$ atomic sites. 
The detailed parametrization of $\gamma_{n,m}$ is given below.

Triptycene is the 3D aromatic molecule
and schematically shown in Fig.~\ref{fig:figure1}(a).  
The black and white circles indicate sp$^2$ and sp$^3$ carbon atoms,
respectively. Each sp$^2$ carbon provides one $\pi$-electron.  
Owing to the central sp$^3$ carbons, three propeller wings extend toward
three directions forming C$_{3v}$ symmetry.  
Since there is no $\pi$-electron on the center of three propellers, the
$\pi$-electrons construct the triangle network in the triptycene. 
Figure~\ref{fig:figure1}(b) shows the top view of triptycene molecule
together with $\pi$-orbitals. 
The electron hoppings within the same wing are defined as $-\gamma_0$, and
those between different wings are defined as $\gamma_1$, where
$\gamma_0>0$ and $\gamma_1>0$.

\begin{figure*}[hbt]
  \begin{center}
    \includegraphics[width=\textwidth]{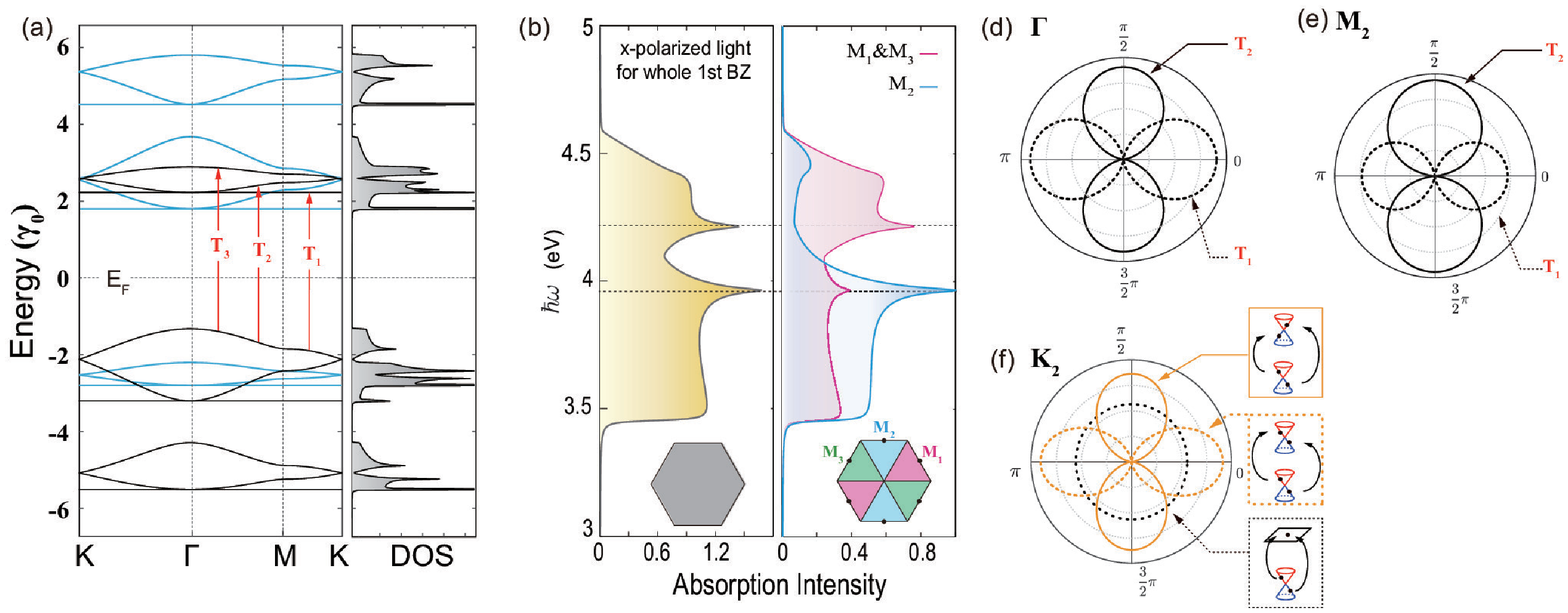}
  \end{center}
  \caption{
    (a) Energy bands structure of zigzag TMM together with corresponding
 DOS on the basis of tight-binding model. Black and cyan lines in the energy band structure indicate bonding
 and anti-bonding states between upper and lower layers,
 respectively. Note that the optical transition between black and cyan
 subbands are prohibited. The representative optical transitions from
 the highest valence subband are named $T_1$, $T_2$ and $T_3$. 
(b) (Left panel) Corresponding absorption spectrum of
 $T_1$, $T_2$ and $T_3$ optical transitions. The $k$-integration is performed within
 the whole 1st BZ. (Right panel) Same for $T_1$, $T_2$ and $T_3$ optical
 transitions, but the $k$-integration is performed in the $1/3$ regions
 of 1st BZ which separately contain $M_1$, $M_2$ and $M_3$ points. Angular dependence of
 absorption intensity at (d) $\Gamma$, (e) $M_2$, and (f) $K_2$ points.
  }
  \label{fig:figure2}
\end{figure*}

As shown in Figs.~\ref{fig:figure1} (c) and (d), 
there are two types of
cross-linked structures of TMMs, i.e., zigzag and armchair TMMs,
respectively. It should be noted that both of TMMs have C$_{6v}$
symmetry.
Here, the magenta  arrows are primitive vectors, given as 
$\bm{a_1} = (a, 0)$ and 
$\bm{a_2} = (-\frac{a}{2}, \frac{\sqrt{3}a}{2})$,
where $a$ is lattice constant: $a=8.92$\AA\ for zigzag TMM and $a=22.17$\AA\ 
for armchair TMM, respectively.
The yellow shaded areas are the unit cells of TMMs.
The zigzag and armchair TMMs have $18$ and $54$ $\pi$-electronic sites
 in their unit cells, respectively.
Since the corresponding reciprocal lattice vectors are given as
$\bm{b_1} = \frac{2\pi}{a}(1, \frac{1}{\sqrt{3}})$ and 
$\bm{b_2} = \frac{2\pi}{a}(0, \frac{2}{\sqrt{3}})$, the 1st BZ for TMMs
becomes the hexagonal shown in Fig.~\ref{fig:figure1}(e).

In zigzag TMMs, triptycene molecules are polymerized by sharing benzene
rings between neighboring molecules.
It can be understood that the $\pi$-electrons form the
network composed of triangle rings and even membered rings as shown
in Fig.~\ref{fig:figure1} (c).
The inplane network structure is resemble to the Kagome lattice,
where triangle and hexagonal-symmetric rings are alternatively spread.
Besides, this Kagome-like network forms the bilayered structure which can
be seen in side-view. The bilayer structure leads to bonding and
anti-bonding molecular orbitals between upper and lower layers. 
As shown in Appendix~\ref{apeA}, the tight-binding model of 2D Kagome lattice 
produces the energy band structures with graphene energy dispersion
and a perfect flat band. 
In actual, the DFT calculations show that zigzag TMMs become
semiconducting and show the energy band structures accompanying several
sets of Kagome-like energy dispersion. For the tight-binding model of
zigzag TMMs, we use 
$-\gamma_0$ for the electron transfer within benzene rings, 
$\gamma_1=\gamma_0/4$ for triangular rings connecting benzene rings.
Throughout this paper, we set $\gamma_0=2.43$eV. 
This parameter set fairly reproduces the energy band structures of
zigzag TMMs obtained DFT calculations.

Figure~\ref{fig:figure1} (d) shows the lattice structure of armchair
TMMs, where triptycene molecules are connected through benzene
molecules with $\sigma$-bondings. Since the benzene molecule bridges two carbon atoms belonging to
the different layers as shown in the side-view of the 
structure, armchair TMM has the rippling structure. 
The DFT study has shown that armchair TMM is energetically stable and
semiconducting. Similar to zigzag TMM, owing to the presence of
triangle rings, armchair TMMs also has Kagome-like energy band
structures. For the tight-binding model of armchair TMM, we use
$-\gamma_0$ for the electron transfer within benzene rings, 
$\gamma_1=\gamma_0/4$ for triangular rings connecting benzene rings.
It is also known that the bridging benzene rings are tilted
with the angle of $\frac{\pi}{6}$ from the vertical plane to the armchair membrane 
owing to the conformation. 
Thus, the transfer integral for the bridging bonds is taken as
$-\gamma_2(=-\gamma_0\cos\theta)$ with $\theta=\frac{\pi}{6}$.
This parameter set fairly reproduces the energy band structures of
armchair TMMs obtained DFT calculations.~\cite{Fujii2}

Let us briefly make the overview of theoretical framework to study the optical
properties of TMMs under the linearly and circularly polarized light
irradiation. 
The light absorption coefficient of solid is described as
\begin{equation*}
 \alpha(\omega) = \frac{\omega}{cn} \rm{Im}[\varepsilon{(\omega)}],
\end{equation*}
where $\varepsilon(\omega)$ is complex dielectric function.
Here $\omega$ is frequency of irradiation light and $n$ is refractive index. 
In addition, dielectric function $\varepsilon(\omega)$ can be related to
the dynamical conductivity $\sigma(\omega)$ by
\begin{equation*}
  \varepsilon(\omega) = 1 + i \frac{4\pi}{\omega} \sigma(\omega).
\end{equation*}
\begin{figure*}
  \begin{center}
    \includegraphics[width=\textwidth]{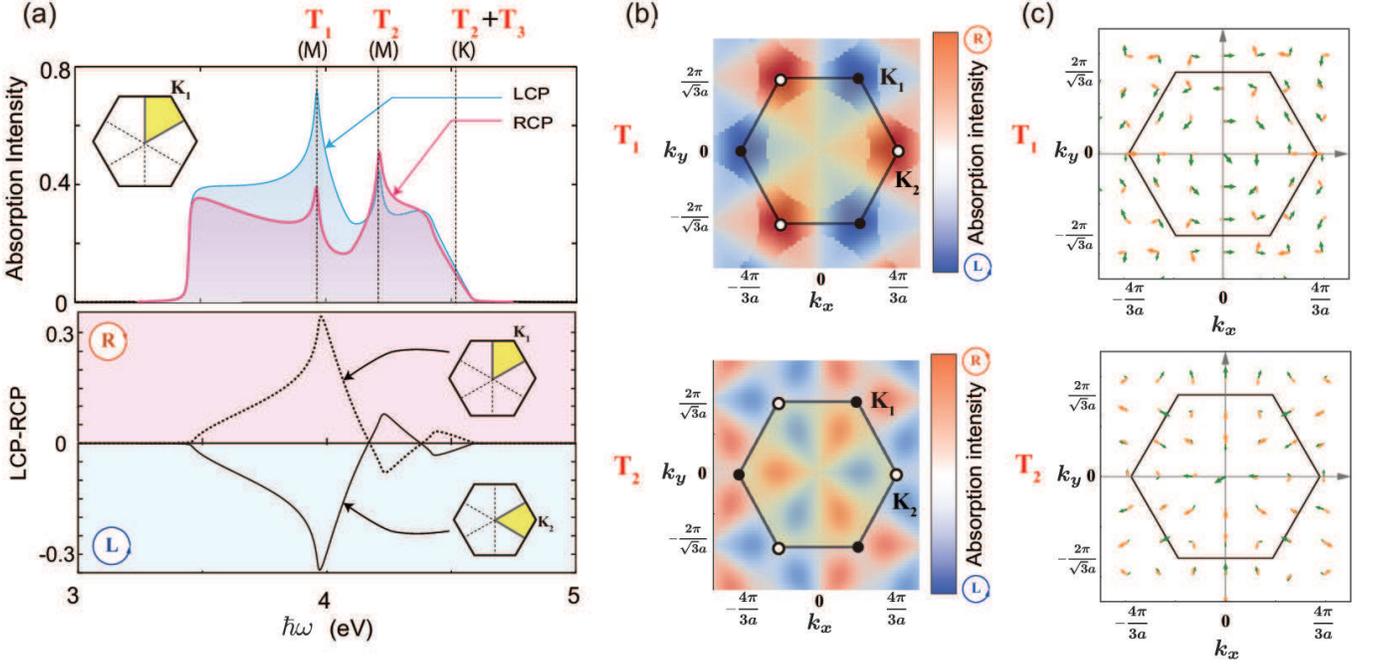}
  \end{center}
  \caption{(a) (Upper panel) Optical absorption spectrum of zigzag TMM under the
 circularly polarized light irradiation. 
The three dashed lines indicate the representative optical transition
 caused by $T_1$ at $M$ point, $T_2$ at $M$ point and $T_2+T_3$ at $K$
 point from left. 
The integration is
 evaluated within the $1/6$ region of 1st BZ containing $K_2$ point.
(Lower panel) The difference of optical absorption between LCP and
 RCP. Dashed and solid lines correspond to the cases of $k$-integration
 for the $1/6$ regions of BZ
 containing $K_1$ and $K_2$, respectively. 
 Since TMM preserves both time-reversal and inversion
symmetries, $K_1$ and $K_2$ states polarize oppositely, i.e., no net valley polarization. 
(b)
Momentum space mapping of absorption intensity difference between LCP
 and RCP, i.e.,  $\Delta\alpha(\bm{k})$ for (upper panel) $T_1$ and (lower panel) $T_2$ transitions. 
(c) The distribution of dipole vector in the momentum space for (upper
 panel) $T_1$ and (lower panel) $T_2$ transitions
under the
 circularly polarized light irradiation. 
The green and orange arrows indicate real and imaginary parts of the
 dipole vector, respectively. 
  }
  \label{fig:figure3}
\end{figure*}

Since the effect of external field is treated within first-order
perturbation, 
the dynamical conductivity can be evaluated through Kubo formula,~\cite{kubo1957jpsj,mahan} 
i.e., 
\begin{align*}
\sigma(\omega) & = \frac{\hbar}{iS} \sum_{\bm{k}} \sum_{i, f} \frac{f{(E_{\bm{k}}^{(f)})} -
  f{(E_{\bm{k}}^{(i)})}} {E_{\bm{k}}^{(f)} - E_{\bm{k}}^{(i)}} 
\frac{\left| \bm{e} \cdot\bra{\psi_{\bm{k}}^{(f)}} \nabla_{\bm{r}}
 \ket{\psi_{\bm{k}}^{(i)}} \right|^2}{E_{\bm{k}}^{(f)} -
 E_{\bm{k}}^{(i)} - \hbar \omega +i \eta}.\\
& \eqqcolon \frac{1}{S} \sum_{\bm{k}} \sum_{i, f} \tilde{\sigma}_{f,i}(\bm{k},\omega)
\end{align*}
$S$ is the area of system and $E_{\bm{k}}^{(i)}$ and $E_{\bm{k}}^{(f)}$
 indicate the eigenenergies for initial and final states
 for the inter-band optical transition, respectively. 
$\psi_{\bm{k}}^{(i)}$ and $\psi_{\bm{k}}^{(f)}$ are corresponding
 wavefunctions obtained from tight-binding model. 
The $\bm{k}$-summation is performed within the 1st BZ.
$f{(E^{(i)}_{\bm{k}})}$ is the Fermi-Dirac distribution function for the state of
 energy $E^{(i)}_{\bm{k}}$. 
$\eta$ is an infinitesimally small real number.
Also, $\bm{e}=(e_x, e_y)$ is the polarization vector of incident light
 and
$\bra{\psi_{\bm{k}}^{(f)}} \nabla_{\bm{r}} \ket{\psi_{\bm{k}}^{(i)}}$ represents the
 transition dipole vector, where $\nabla_{\bm{r}}=(\partial/\partial x,
 \partial/\partial y)$.
For later use, we have also defined the integrand of $\sigma(\omega)$ as
$\tilde{\sigma}_{f,i}(\bm{k},\omega)$, which gives the momentum resolved
 aborption intensity from $\psi_{\bm{k}}^{(i)}$ to
 $\psi_{\bm{k}}^{(f)}$, i.e., $\alpha_{f,i}(\bm{k},\omega)$.
This will be used for momentum space mapping of aborption intensity.

The transition dipole vector is evaluated as the
expectation value of group velocity,~\cite{PhysRevB.47.15500} i.e., 
\begin{align*}
  \bra{\psi_{\bm{k}}^{(f)}} \nabla_{\bm{r}} \ket{\psi_{\bm{k}}^{(i)}} 
    =i\frac{m}{\hbar} \bra{\psi_{\bm{k}}^{(f)}} \frac{\partial \hat{H}}{\partial \bm{k}} \ket{\psi_{\bm{k}}^{(i)}}. 
\end{align*}
The inner product between the polarization vector and the transition
dipole vector leads to 
\begin{align*}
&\bm{e} \cdot \bra{\psi_{\bm{k}}^{(f)}} \nabla_{\bm{r}} \ket{\psi_{\bm{k}}^{(i)}} \\
  &=i\frac{m}{\hbar} \left(  e_x \bra{\psi_{\bm{k}}^{(f)}} \frac{\partial \hat{H}}{\partial \bm{k_x}} \ket{\psi_{\bm{k}}^{(i)}} + e_y \bra{\psi_{\bm{k}}^{(f)}} \frac{\partial \hat{H}}{\partial \bm{k_y}} \ket{\psi_{\bm{k}}^{(i)}} \right). 
\end{align*}

The polarization of light can be incorporated through the Jones
vectors.~\cite{jones1941} For linearly polarized light, it is given as 
\begin{displaymath}
  \bm{e}_{\rm{Linear}} =\left( \begin{array}{c} \cos{\phi} \\  \sin{\phi} \\\end{array} \right),
\end{displaymath}
where $\phi$ is direction of electric field of incident light
measured from $x$-axis. Meanwhile, for right-handed circularly polarized
(RCP) light irradiation, we use 
\begin{align*}
  \bm{e}_{\rm{RCP}} = \frac{1}{\sqrt{2}} \left( \begin{array}{c} 1 \\
					  -i \\\end{array} \right),
\end{align*}
and for left-handed circularly polarized (LCP) light, we use
\begin{align*}
  \bm{e}_{\rm{LCP}} = \frac{1}{\sqrt{2}} \left( \begin{array}{c} 1 \\ i \\\end{array} \right).
\end{align*}
$\bm{e}_{\rm{RCP}}$ and $\bm{e}_{\rm{LCP}}$ satisfy the orthogonality. 

\section{Optical properties of zigzag TMM}
In this section, we consider the optical properties of zigzag TMM under
linearly and circularly polarized light
irradiation.
Figure~\ref{fig:figure2} (a) shows the energy band
structure of zigzag TMM together with the corresponding density of
states (DOS) on the basis of tight-binding model.
The system is semiconducting with the direct band gap. 
Since zigzag TMM has C$_{6v}$ symmetry same as the 2D Kagome lattice, 
several Dirac cones appear at $K$-point. 
Simultaneously, several perfect flat bands appear owing to the nature of Kagome lattice. 
If we say a set of graphene-like bands with a flat band as a Kagome-like
energy band, as we have expected, six set of Kagome-like energy band structures are obtained. 
Since zigzag TMM has the bilayered structure, we can distinguish the
energy subbands into bonding (black line spectrum) and anti-bonding (cyan
line spectrum) subbands across the upper and lower layers.
It is noted that optical transitions occur between same types of states,
i.e., the transition from bonding (anti-bonding) to anti-bonding
(bonding) states is forbidden since the parity of the wavefunction is reversed with respect to the $xy$-plane.

\subsubsection{Linearly Polarized Light}
Let us closely inspect the optical properties of zigzag TMM under
linearly polarized light irradiation. To study the details of optical
selection rules, we focus on the optical transition from the highest
valence subband. The representative optical transitions are named $T_1$,
$T_2$ and $T_3$ indicated by red arrows in
Fig.~\ref{fig:figure2}(a). Note that the optical transition to
cyan-colored subband is prohibited. 
The left panel of Fig.~\ref{fig:figure2}(b) shows
the incident energy dependence of absorption intensity under linearly
polarized light irradiation. Only $T_1$, $T_2$ and $T_3$ transitions
contribute optical absorption in this energy region, and
$k$-integration is performed in the whole 1st BZ.

The absorption spectrum shows two intensive peaks at $3.96$ and
$4.21$ eV, which originate from the divergent joint density of states
(JDOS) owing to the saddle points of energy band structure at $M$
points. The first peak mainly arises from the optical transition
$T_1$, and the second one arises from $T_2$.

It is noted that three non-equivalent $M$ points contribute differently
to these peaks of the optical absorption spectrum. 
The right panel of Fig.~\ref{fig:figure2}(b) shows the contributions from 
$M_1$, $M_2$, and $M_3$ points.
To separate the contribution from each $M$ point, 
$\bm{k}$-integration of optical conductivity is performed in the three
divided regions as shown in the inset of right panel
of Fig.~\ref{fig:figure2}(b). 
It is seen that
the $M_2$ point has the larger contribution to the first peak than $M_1$ and $M_3$,
but the smaller contribution to the second peak. Thus, it is possible to make a
polarization among three non-equivalent $M$ points using the
linearly polarized light irradiation. 
\begin{figure*}[hbt]
  \begin{center}
    \includegraphics[width=\textwidth]{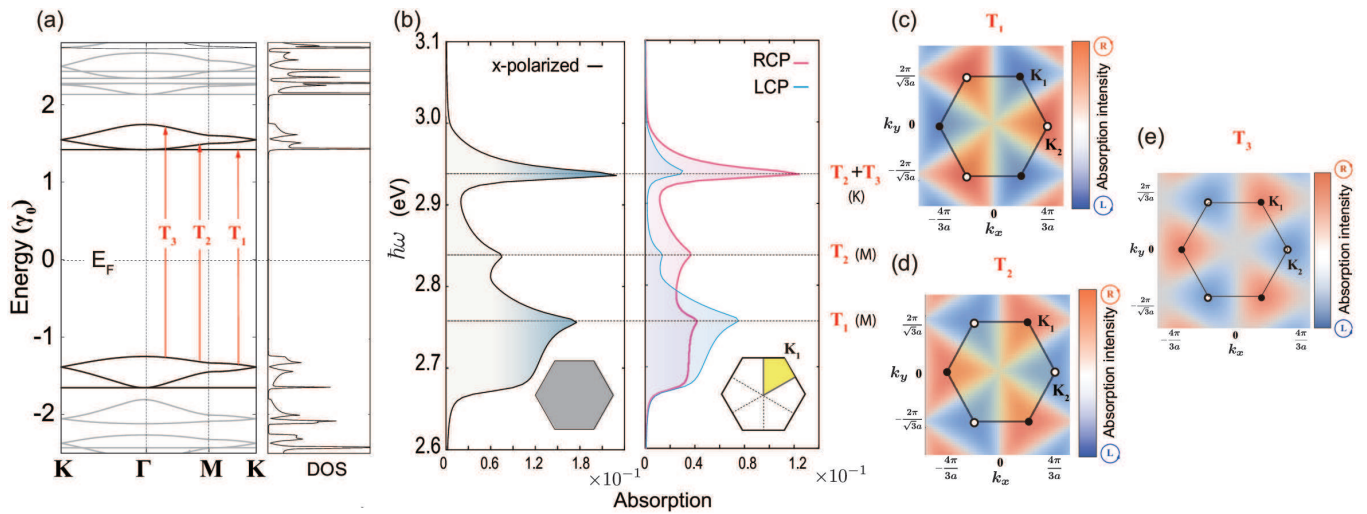}
  \end{center}
  \caption{(a) Energy band structure of armchair TMM together with
 DOS on the basis of tight-binding model. The optical transitions from the highest valence subband to lowest
 three conduction subbands are named $T_1$, $T_2$ and $T_3$. 
 (b) Optical absorption spectrum of armchair TMM under (left panel)
 linearly and (right panel) circularly irradiation. For linearly
 irradiation, $k$-integration is performed for whole 1st BZ.
There are three pronounced peaks at $\hbar\omega=2.76$, $2.84$, and
 $2.94$ eV. At right side of the panel, it is noted that dominant optical
 transition and the high symmetric points where the optical transitions occur.
 For circularly irradiation, $k$-integration is performed for
 the $1/6$ region of 1st BZ containing $K_1$.
Momentum space mapping of absorption intensity difference between RCP
 and LCP, i.e.,  $\Delta\alpha(\bm{k})$ for 
(c) $T_1$, (d) $T_2$ and (e) $T_3$ transitions.}
  \label{fig:figure4}
\end{figure*}

\subsubsection{Polarization Angle Dependence at High Symmetric $k$-Points}
Next, we shall discuss the polarization angle dependence of
optical absorption intensities at $\Gamma$, $M$, $K$ points using group theory.~\cite{dresselhausgrouptheory}
Since the TMMs have similar crystal symmetry with 2D Kagome lattice, 
the polarization angle dependence is quite analogous to the case of
2D Kagome lattice discussed in Appendix~\ref{apeA} in
detail. 
 
\noindent
{\bf [1] $\Gamma$ point:} 
The $\Gamma$ point has C$_{6v}$ symmetry. However, in Kagome lattice 
and TMMs,
the optical selection rules are determined by C$_{3v}$ symmetry, 
because of the existence of the triangle unit in the lattice. Thus, 
optical transition occurs between non-degenerate $A_1$ state and doubly
degenerate $E$ states. Therefore, only $T_1$ and $T_2$ transitions are
optically active, but $T_3$ is prohibited. 

Figure~\ref{fig:figure2}(d) shows the polar angle dependence of light
absorption at the $\Gamma$ point for $T_1$ and $T_2$. 
Since the wavefunction of zigzag TMM (not shown) has the same symmetry as that of 2D
Kagome lattice, the $T_1$ and $T_2$ have the polarization angle
dependence on $\cos\phi$ and $\sin\phi$, respectively. 

\noindent
{\bf [2] $M$ point:} 
The $M$ points have C$_{2v}$ symmetry, i.e., there is no degeneracy in energy
dispersion at these points. In zigzag TMM, all the wavefunctions at $M$
points are classified into either $A$ or $B$ representation, as similar to
the case of 2D Kagome lattice. The optical transitions occur for $T_1$
and $T_2$. It should be noted that $T_3$ is not allowed because $T_3$
connects the two states with the same symmetry. 
In Fig.~\ref{fig:figure2}(e), the angle dependence of linearly polarized
light at $M_2$ point is shown. 
We can clearly confirm the $T_1$ and $T_2$ have $\cos\phi$ and
$\sin\phi$ dependence, respectively. This is consistent with the
case of 2D Kagome lattice. 
For $M_1$ ($M_3$) point, the polarization angle dependence is 
obtained by shifting the angle as $\phi \rightarrow \phi-\pi/3$
($\phi+\pi/3$). 

\noindent
{\bf [3] ${K}$ point:} 
The $K$ points have C$_{3v}$ symmetry. Similar to 2D Kagome lattice,
non-degenerate $A_1$ states and doubly degenerates $E$ states appear at
$K$ points in zigzag TMM. The flat band state corresponds to $A_1$, and 
the Dirac points correspond to $E$ states. 
Thus, the optical transitions are allowed between $A_1$ and $E$ states,
i.e, $T_1$, and between $E$ states, i.e., $T_2$ and $T_3$.
However, it is noted that the optical absorption is relatively weak
compared with those at $\Gamma$ and $M$ points, owing to the smaller JDOS
of Dirac cones in zigzag TMM.

The angle dependence at $K$ points is shown in
Fig.~\ref{fig:figure2}(f). It is noted that the optical transition 
of $T_1$ is isotropic, which is not seen in $\Gamma$ and $M$ points.
It is equivalent to the polarization dependence of 2D Kagome
lattice, see Appendix~\ref{apeA}. 

Furthermore, in TMMs, the unique optical transitions occur
between Dirac cones, which are absent in simple 2D Kagome lattice.
Similar to graphene, the Dirac cones are expected to have the 
helicity. In general, the upper and lower cones have opposite
helicities even at the same $K$ point. The red and blue Dirac cones in
Fig.~\ref{fig:figure2}(f) indicate the opposite helicity. It should be
noted that the optical transitions between the same (different)
helicities have the polarization angle dependence of $\sin\phi$
($\cos\phi$). Thus, the polarization dependence of absorption spectrum
at $K$ points has different origin from those at the $\Gamma$ and $M$ points. 

\subsubsection{Circularly Polarized Light}
Since zigzag TMM has the valley structures in the energy band structure, 
the circularly polarized light irradiation can selectively excite the electrons belonging to 
either ${K_1}$ or ${K_2}$ point depending on the direction of
polarization.
Note that the present system preserves both time-reversal and inversion
symmetries, no net valley polarization occurs. 
The upper panel of Fig.~\ref{fig:figure3}(a) is the absorption spectrum
under the circularly polarized light irradiation,
where the $k$-integration has been performed within the $1/6$ region of BZ
containing $K_2$ point, see inset of Fig.~\ref{fig:figure3}(a).
It can be noticed that there are two pronounced peaks around $\hbar\omega=3.96$ and $4.21$ eV,
where only the first peak shows rather large difference between LCP and RCP. 
The first and second peaks correspond to the optical transition of $T_1$ and $T_2$ at $M$ point, respectively. 
The optical transition of $T_1$ is mainly dominated by the electronic excitation at $M$ and $K$ points.
However, $T_2$ is mostly dominated by electronic excitation at $M$ points.
The lower panel of Fig.~\ref{fig:figure3}(a) is the difference of optical absorption between LCP and
 RCP. Since TMM preserves both time-reversal and inversion
symmetries, $K_1$ and $K_2$ states polarize oppositely, i.e., no net valley polarization.

Figure~\ref{fig:figure3}(b) shows momentum space mappings of
absorption intensity differences between LCP and RCP, $\Delta\alpha(\bm{k})$, for the optical transitions $T_1$ and $T_2$.
$\Delta\alpha(\bm{k})$ is defined as
$\Delta\alpha(\bm{k})\coloneqq \alpha^{LCP}_{(f,i)}(\bm{k},\omega)-\alpha^{RCP}_{(f,i)}(\bm{k},\omega)$,
where $(f,i)$ and $\omega$ are chosen to satisfy the condition of
specific optical transition.
Red (blue) region indicates strong absorption for RCP (LCP). 
The strong valley selective excitation by circularly
polarized light irradiation can be observed for the optical transition
$T_1$. However, such valley selective excitation becomes weak for $T_2$. 
This can be understood by inspecting the momentum space mapping of
the dipole vector, $\bra{\psi_{\bm{k}}^{(f)}} \nabla_{\bm{r}}\ket{\psi_{\bm{k}}^{(i)}}$
as shown in Fig.~\ref{fig:figure3}(c),
where green and orange arrows indicate real and imaginary parts of the
dipole vector, respectively. 
As can be seen, the real and imaginary parts are orthogonal near the
$K_1$ and $K_2$ points for $T_1$, resulting in 
the valley selective excitation.~\cite{JPSJ.87.024710,PhysRevB.97.195444}
However, the real and imaginary parts of dipole vector
becomes parallel in the transition $T_2$, i.e. very weak valley selective
excitation.
Note that the momentum mapping of absorption intensity for $T_3$ is not
shown here, because of that $T_3$ is optically prohibited 
except the vicinity of $K$ points.
\begin{figure}[hbt]
  \begin{center}
    \includegraphics[width=0.45\textwidth]{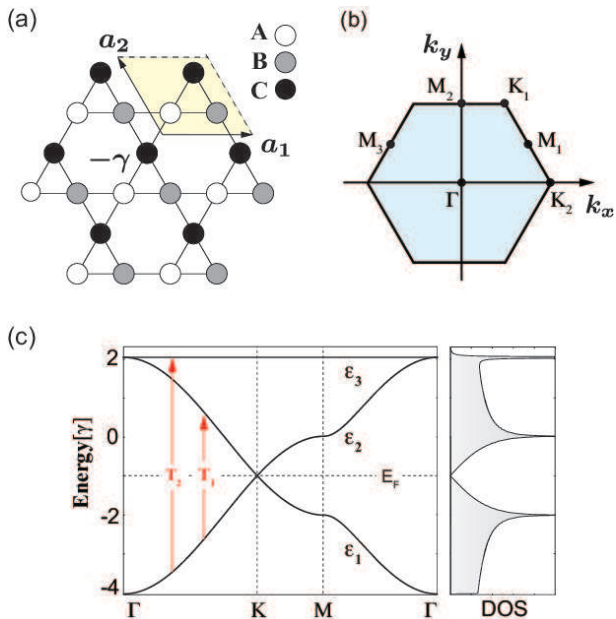}
  \end{center}
  \caption{(a) Schematic of 2D Kagom{e} lattice. The yellow area
 represents the unit cell and the primitive vectors are $\bm{a_1} = (a,
 0)$, $\bm{a_2} = (-\frac{a}{2}, \frac{\sqrt{3}a}{2})$. $-\gamma$ is the
 hopping parameter between the nearest-neighbor sites. (b) 1st BZ. (c)
 Energy band structure and DOS. The optical transition from $\epsilon_1$
 to $\epsilon_2$ ($\epsilon_3$) is defined as $T_1$ ($T_2$).}
  \label{fig:figure5}
\end{figure}

\section{Optical properties of armchair TMM}
In this section, we briefly discuss the optical properties of armchair
TMM. For armchair TMM, we can apply the similar optical selection rules found in
zigzag TMM. However, owing to the less dispersive energy band structures 
of armchair TMM, rather clear valley selective optical excitation can be
observed. 

Figure~\ref{fig:figure4}(a) shows the energy band structure of armchair
TMM together with the corresponding DOS on the basis of tight-binding model.
Here the titled angle is set to $\frac{\pi}{6}$. 
Similarly, several sets of Kagome bands are obtained. Since 
the unit cell contains $54$ atomic sites, only subbands and DOS near the 
Fermi energy are shown. Owing to the rippling structure of armchair
TMMs, we cannot decompose the energy subbands into bonding and
anti-bonding states.
Similar to the case of zigzag TMM, let us focus on the optical 
transition from the highest valence subband to three lowest conduction
subbands indicated as $T_1$, $T_2$, and $T_3$ in Fig.~\ref{fig:figure4}(a).
Note that no valley polarization occurs, because both time-reversal and inversion symmetries are 
preserved.

In armchair TMM, the absorption spectrum of these transitions
has the two strong peaks around the $2.76$eV and $2.94$eV
in addition to one weak peak around the $2.84$eV as shown in the
left panel of Fig.~\ref{fig:figure4}(b).
Here, $k$-integration is performed in the whole 1st BZ.
The first strong peak and
second weak peak indicate the $T_1$ and $T_2$ transitions at $M$ point,
respectively. However, the third strong peak indicates $T_2$ and $T_3$
transitions at $K$ point. 
In armchair TMM, 
the slope of energy dispersion for Dirac cones becomes smaller than that
of zigzag TMM, leading to faster increase of DOS near the Dirac points. 
This fact induces rather strong valley selective optical absorption at
$K$ point for $\hbar\omega=2.94$eV. 
Thus, armchair TMM can generate the valley selective optical excitation more
clearly, as shown in the right panel of
Fig.~\ref{fig:figure4}(b). 
Here, $k$-integration is performed in the $1/6$ region of 1st BZ containing
$K_1$ point, see inset of Fig.~\ref{fig:figure4}(b). 
Certainly, the momentum space mappings of the
absorption spectrum differences $\Delta\alpha(\bm{k})$ for $T_1$, $T_2$
and $T_3$ transitions clearly indicate the selective
excitation around the $K_1$ and $K_2$ points as shown in
Figs.~\ref{fig:figure4}(c), (d) and (e).

\section{Summary}
In summary, we have theoretically investigated the optical properties of
TMMs under the linearly and circularly polarized light irradiation.
To analyze the optical properties of TMMs, we have constructed the
tight-binding model that faithfully reproduces the energy band structures
obtained by first-principles calculations.
On the basis of the tight-binding model,
we have numerically evaluated the optical absorption
intensity and valley selective optical excitation using the Kubo
formula. This approach reduces significantly the computational cost,
since the TMMs contain the large number of atoms in their unit cells. 
It is found that absorption intensity crucially depends on both light
 polarization angle and the momentum of optically excited electrons.
It has been also confirmed that the circularly polarized light irradiation can
selectively excite the electrons in either $K$ or $K^\prime$ point. 
Besides the circularly polarized light irradiation, the use of second optical harmonics~\cite{PhysRevB.90.201402} 
is another way to generate the valley polarization in TMMs, which will be
studied in future. 
Thus, TMMs are considered to be the good platform for the valleytronics
applications. 
In addition, we have analyzed the selection rules of 
TMMs using group theory, which shows very similar optical selection
rules to that of 2D Kagome lattice system.  
From the analysis, we have determined the absorption
spectrum and polarization-dependent transition at high symmetric points
in 1st BZ. Our works will serve for designing further TMMs and analyze
the experimental data of the optical absorption spectrum of TMMs.

K.W. acknowledges the financial support from Masuya Memorial Research Foundation of
Fundamental Research. This work was supported by JSPS KAKENHI Grant
No. JP18H01154, and JST CREST Grant No. JPMJCR19T1.

\appendix
\section{Selection Rule of 2D Kagome Lattice}\label{apeA}
Here we consider the electronic states of 2D Kagome lattice
and summarize the optical selection rules on the basis of
nearest-neighbor tight-binding model. Figure~\ref{fig:figure5} (a) shows
schematic of 2D Kagome lattice. The yellow shaded area is the unit cell,
which contains three non-equivalent atomic sites called $A$, $B$ and
$C$. The primitive vectors are given as 
$\bm{a_1} = (a, 0)$ and 
$\bm{a_2} = (-\frac{a}{2}, \frac{\sqrt{3}a}{2})$,
where $a$ is the lattice constant.
Here, we assume that each site has a single electronic orbital and
electron hopping parameter between nearest-neighbor sites is $-\gamma$.

The Schr\"odinger equation for 2D Kagome lattice is given as
\begin{align*}
 \hat{H}_{\bm{k}}\psi_{\bm{k}} = \epsilon_{\bm{k}}\psi_{\bm{k}},
\end{align*}
where $\psi_{\bm{k}}=\left(A_{\bm{k}}, B_{\bm{k}}, C_{\bm{k}}\right)$
is the wavefunction representing the amplitude at $A$, $B$ and $C$ subblattice sites in the unit cell,
respectively. $\epsilon_{\bm{k}}$ is the energy eigenvalue.
The Hamiltonian is given by
\begin{equation*}
  \hat{H}_{\bm{k}} = -\gamma\left(
    \begin{array}{ccc}
      0& 1+e^{iK_1} & 1+e^{-iK_3} \\
     1+e^{-iK_1} & 0 & 1+e^{iK_2} \\
      1+e^{iK_3} & 1+e^{-iK_2} & 0
    \end{array}
  \right).
\end{equation*}
Here, we have defined 
\begin{math}
	K_\nu=\bm{k}\cdot\bm{a}_{\nu}/2.
\end{math}
with $\nu = (1,2,3)$, $\bm{k}=(k_x,k_y)$ and $\bm{a_3}=-(\bm{a_1}+\bm{a_2})$.
Since the reciprocal lattice vectors are given as
$\bm{b_1} = \frac{2\pi}{a}(1, \frac{1}{\sqrt{3}})$ and 
$\bm{b_2} = \frac{2\pi}{a}(0, \frac{2}{\sqrt{3}})$, the 1st BZ of 2D 
Kagome lattice becomes hexagonal shown in Fig.~\ref{fig:figure5}(b).
The energy eigenvalues are given as
\begin{math}
\epsilon_{\bm{k}}/\gamma=2,
-1\pm\sqrt{3+2\sum_{\nu=1}^3\cos\left({K_\nu}\right)}.
\end{math}
It should be noted that the form
$\pm\sqrt{3+2\sum_{\nu=1}^3\cos\left({K_\nu}\right)}$ is completely
identical with the form of energy band dispersion of nearest-neighbor
tight binding model for $\pi$-electrons of graphene.

Figure~\ref{fig:figure5}(c) shows energy band structure and corresponding density of states of 2D Kagome lattice.
There are three subbands in this system, the two lowest subbands have
the identical structure with that of 2D honeycomb lattice, i.e.,
graphene. However, there is the perfect flat band over the 1st BZ at
$\epsilon=2\gamma$. Hereafter, we call the energy subband $\epsilon_1$,
$\epsilon_2$ and $\epsilon_3$ from lowest one to highest one, 
and corresponding wavefunctions $\psi_1$, $\psi_2$ and $\psi_3$, respectively. 
We also define that the optical transition from $\epsilon_1$ to $\epsilon_2$ ($\epsilon_3$) as $T_1$ ($T_2$). 
\begin{table}[htb]
  \begin{center}
    \caption{Character table of $C_{6v}$}
    \begin{tabular}{l|c c c c c c|l} 
      $C_{6v}$ & $E$ & $2C_6(z)$ & $2C_3(z)$ & $C_2(z)$ & $3\sigma_{v}$ & $3\sigma_{d}$ &  \\ \hline 
      $A_{1}$ &  1 &  1 &  1 &  1 &  1 &  1 & $z$      \\
      $A_{2}$ &  1 &  1 &  1 &  1 & -1 & -1 &          \\
      $B_{1}$ &  1 & -1 &  1 & -1 &  1 & -1 &          \\ 
      $B_{2}$ &  1 & -1 &  1 & -1 & -1 &  1 &          \\
      $E_{1}$ &  2 &  1 & -1 & -2 &  0 &  0 & $(x, y)$ \\
      $E_{2}$ &  2 & -1 & -1 &  2 &  0 &  0 &          \\ 
    \end{tabular}
    \label{C6v}
  \end{center}
\end{table}
\begin{table}[htb]
  \begin{center}
    \caption{Character table of $C_{3v}$}
    \begin{tabular}{l|c c c |l} 
      $C_{3v}$ & $E$ & $2C_3(z)$ & $3\sigma_{v}$ &  \\ \hline 
      $A_{1}$ & 1 &  1 &  1 & $z$     \\
      $A_{2}$ & 1 &  1 & -1 &       \\ 
      $E$    & 2 & -1 &  0  & $(x, y)$\\
    \end{tabular}
    \label{C3v}
  \end{center}
\end{table}
\begin{figure*}[hbt]
  \begin{center}
    \includegraphics[width=1\textwidth]{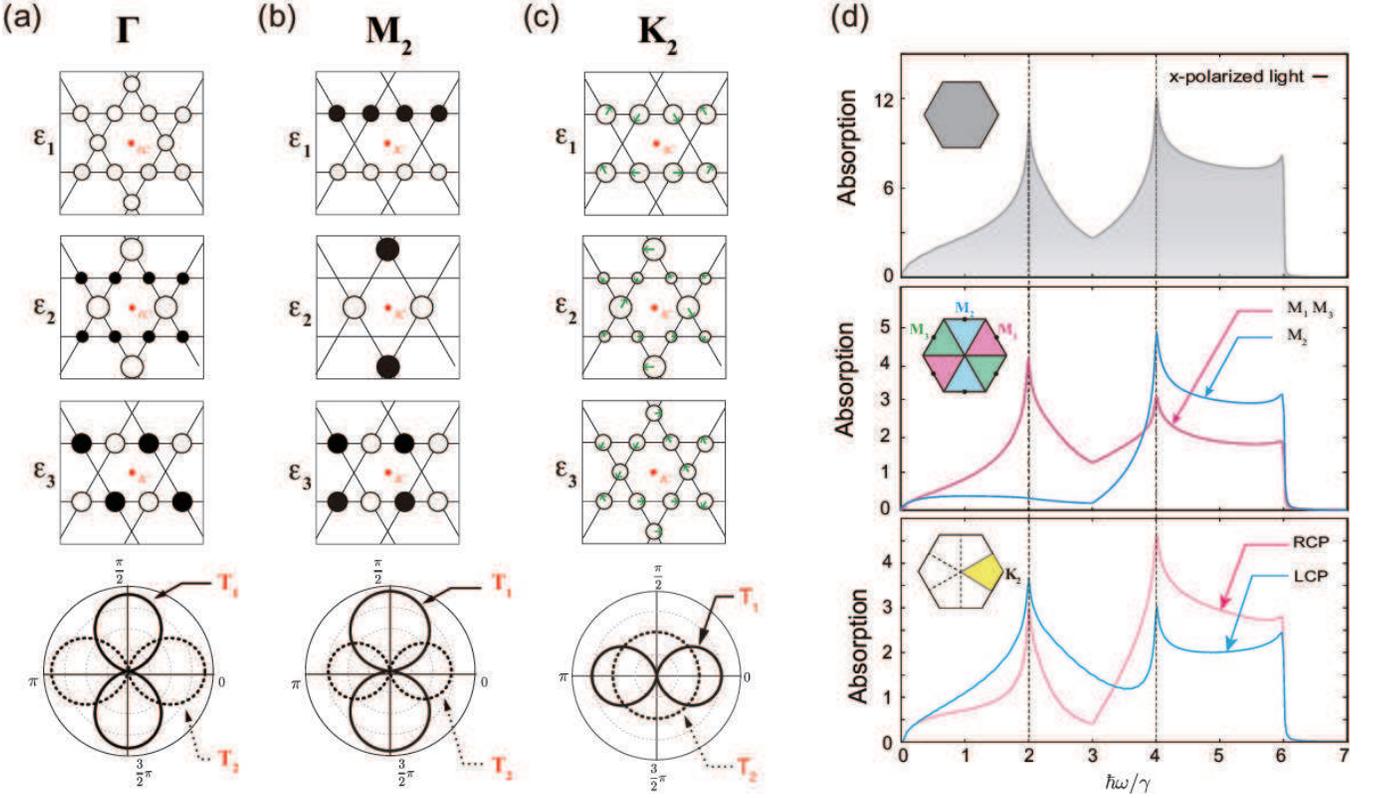}
  \end{center}
  \caption{
The real space distribution of wave functions for 2D Kagome lattice and
 (bottom panel) corresponding angle dependence of optical absorption at high symmetric points:
 (a) $\Gamma$ and  (b) $M_2$.
Black and white circles indicate the sign of wave functions, and the
 radius of circles indicates the amplitude of wave functions. 
Red dots indicate the inversion center (IC) of Kagome lattice. 
(c) Same for $K_2$ point. Here, the argument of complex wave function is indicated by green arrows.
(d) Absorption intensity and polarization of 2D
 Kagome lattice, where $k$-integration for optical conductivity is
 performed (upper panel) for the whole 1st BZ, (middle panel) for $1/3$ regions of 1st
 BZ containing either $M_1$, $M_2$ or $M_3$ point, (lower panel) for $1/6$
 region of 1st BZ containing $K_2$ point.
  }
  \label{fig:figure6}
\end{figure*}

Let us discuss the optical selection rules and polarization angle dependence of
2D Kagome lattice at high symmetric $\bm{k}$-points, i.e., $\Gamma=(0,0)$, 
$K_2=\frac{2\pi}{a}(\frac{2}{3}, 0)$, and
$M_1=\frac{2\pi}{a}(0, \frac{1}{\sqrt{3}})$. 

\noindent
{\bf [1] $\Gamma$ point}: 
The $\Gamma$ point has C$_{6v}$ symmetry, which obeys the character
table of Table~\ref{C6v}.
At $\Gamma$ point, the eigenenergies are $\epsilon_1=-4\gamma$, and
$\epsilon_2=\epsilon_3=2\gamma$. The wavefunction is analytically given as
\begin{align*}
\psi^{\Gamma}_{1}& = \frac{1}{\sqrt{3}} \left(1, 1, 1 \right),\\ 
\psi^{\Gamma}_{2}& = \sqrt{\frac{2}{3}} \left(-\frac{1}{2}, -\frac{1}{2}, 1 \right),\\
\psi^{\Gamma}_{3}& = \frac{1}{\sqrt{2}} \left(1, -1, 0 \right),
\end{align*}
which are schematically drawn in Fig~\ref{fig:figure6}(a).
According to the character table of C$_{6v}$, 
$\psi_{1}$ is $A_1$ representation, and the degenerate states of $\psi_{2}$ and $\psi_{3}$ 
are $E_2$ representation. Since the polarization vectors belong to $E_1$ representation, 
the tensor product $E_1\otimes A_1=E_1$ 
indicates that the optical transition to $E_2$ is not allowed.
However, it should be noted that 2D Kagome lattice contains
C$_{3v}$ symmetry if we take the triangle unit as the symmetry center.   
According to the character table of C$_{3v}$ (see Table~\ref{C3v}), 
$\psi_{1}$ is $A_1$ representation, and the degenerate states of $\psi_{2}$ and $\psi_{3}$ 
are $E$ representation. Since in C$_{3v}$ the tensor product is given as
$E\otimes A_1=E$, both of optical transitions $T_1$ and $T_2$ are active.
Furthermore, the basis functions for $\psi_1$, $\psi_2$ and $\psi_3$
states are given as $z$, $y$, $x$, respectively. The optical transition
$T_1$ and $T_2$ have $x$ and $y$ polarization, respectively. 

Since the wavefunctions are analytically obtained, we can analytically
evaluate expectation values of the optical dipole vectors. 
For linear polarization, we obtain 
\begin{align*}
 \bm{e}\cdot \langle\psi_2| \nabla_{\bm{k}} H|\psi_1 \rangle & = 
-i\frac{3\sqrt{3}a}{2\sqrt{2}}\sin\phi,\\
 \bm{e}\cdot \langle\psi_3| \nabla_{\bm{k}} H|\psi_1 \rangle & = 
i\frac{3a}{\sqrt{6}}\cos\phi.
\end{align*}
This polarization angle dependence is consistent with the numerical calculations
as shown in the bottom panel of Fig.~\ref{fig:figure6}(a).
For circularly polarization, we obtain
\begin{align*}
 \left| \bm{e}_{LCP}\cdot \langle\psi_j| \nabla_{\bm{k}} H|\psi_1 \rangle \right|^2 & = 
 \left| \bm{e}_{RCP}\cdot \langle\psi_j| \nabla_{\bm{k}} H|\psi_1 \rangle  \right|^2,
\end{align*}
where $j=2,3$.
Thus, there is no dependence on direction of circularly polarization at
$\Gamma$ point. 
\begin{table}[htb]
  \begin{center}
    \caption{Character table of $C_{2v}$}
    \begin{tabular}{l|c c c c|l} 
      $C_{2v}$ & $E$ & $C_2$ & $\sigma_{v}(xz)$ & $\sigma_{v^{\prime}}(yz)$ & \\ \hline 
      $A_{1}$ & 1 &  1 &  1 &  1 & $z$\\
      $A_{2}$ & 1 &  1 & -1 & -1 &  \\
      $B_{1}$ & 1 & -1 &  1 & -1 & $x$\\ 
      $B_{2}$ & 1 & -1 & -1 &  1 & $y$\\ 
    \end{tabular}
    \label{C2v}
  \end{center}
\end{table}

\noindent
{\bf [2] $M$ point}: 
The $M$ points have C$_{2v}$ symmetry, which obeys the character
table of Table~\ref{C2v}. The eigenenegies are $\epsilon_1=-2\gamma$, 
$\epsilon_2=0$, and $\epsilon_3=2\gamma$, respectively. 
For example, the wavefunctions at $M_2$ point are given as
\begin{align*}
  \psi^{M_2}_{1} & = \frac{1}{\sqrt{2}}\left(1, 1, 0\right),\\
  \psi^{M_2}_{2} & = \left(0, 0, 1\right), \\
  \psi^{M_2}_{3} & = \frac{1}{\sqrt{2}}\left(1, -1, 0\right),
\end{align*}
which are schematically shown in Fig.~\ref{fig:figure6} (b).
Since C$_{2v}$ is 1D representation, we can define two symmetry axes in real space.
$\psi^{M_2}_{1}$, $\psi^{M_2}_{2}$, and $\psi^{M_1}_{3}$ along the $x$
($y$) direction are $A_1$, $A_1$ and $B_1$ ($B_2$, $A_1$ and $A_1$)
representations, respectively.  
Note that the basis functions for $B_1$ and $B_2$ are $x$ and $y$,
respectively. 
Thus, optical transition $T_1$ ($T_2$) is active with $y$ ($x$) polarization. 
For linear polarization, we can explicitly write the expectation value of dipole vector as
\begin{align*}
 \bm{e}\cdot \langle\psi^{M_2}_2| \nabla_{\bm{k}} H|\psi^{M_2}_1 \rangle & = 
-i\sqrt{\frac{3}{2}}\sin\phi,\\
 \bm{e}\cdot \langle\psi^{M_2}_3| \nabla_{\bm{k}} H|\psi^{M_2}_1 \rangle & = 
-i\cos\phi.
\end{align*}

$M_2$ point has mirror with respect to $k_y$ axis, however, $M_1$ and $M_3$ have
the mirror with respect to $k_y=\frac{1}{\sqrt{3}}k_x$
and $k_y=-\frac{1}{\sqrt{3}}k_x$, respectively. 
Thus, polarization angle dependence for $M_1$ and
$M_3$ can be obtained by replacing the result of $M_2$ as $\phi\rightarrow\phi-\pi/3$ and $\phi\rightarrow\phi+\pi/3$, respectively.
Thus, by tuning the angle of linear polarization, the electrons can be optically excited 
either $M_1$, $M_2$ or $M_3$ point selectively. 

\noindent
{\bf [3] $K_2$ point}: 
The $K_1$ and $K_2$ points have C$_{3v}$ symmetry. The eigenenergies are
$\epsilon_1=\epsilon_2=-\gamma$ and $\epsilon_3=2\gamma$. 
The wavefunction at $K_2$ point is given as
\begin{align*}
\psi^{K_2}_{1} &= \frac{1}{\sqrt{2}}(-\omega^{-1}, 0 ,1),\\
\psi^{K_2}_{2} &= \frac{1}{\sqrt{2}}(1-2\omega, 2 ,-\omega^{-1}),\\
\psi^{K_2}_{3} &= \frac{1}{\sqrt{3}}(\omega^{-1}, \omega, 1),
\end{align*}
where $\omega=\exp\left(i\frac{2\pi}{3}\right)$. 
$\psi_{3}$ has $A_1$ representation, and degenerate states of
$\psi_{1}$ and $\psi_{2}$ have $E$ representation. For degenerate
$\psi_{1}$ and $\psi_{2}$ states, we made the orthonormalization. 
Since the tensor product leads to $E\otimes E = A_1$, the optical 
transition from Dirac cone to the flat band is active. 

For linear polarization, we can explicitly write the expectation value of dipole vector as
\begin{align*}
 \bm{e}\cdot \langle\psi^{K_2}_3| \nabla_{\bm{k}} H|\psi^{K_2}_1 \rangle & = 
-i\frac{\sqrt{3}}{2\sqrt{2}}\omega^{-1}\exp\left(i\phi\right),\\
 \bm{e}\cdot \langle\psi^{K_2}_3| \nabla_{\bm{k}} H|\psi^{K_2}_2 \rangle & = 
\frac{\sqrt{3}}{2\sqrt{2}}\omega^{-1}\exp\left(-i\phi\right).
\end{align*}
The polarization angle works only for the phase factor, i.e., no polarization angle dependence.
This behavior is shown in the bottom panel of Fig.~\ref{fig:figure6}(c).
For circularly polarized light, the expectation values of dipole vector are written as
\begin{align*}
 \bm{e}_{RCP}\cdot \langle\psi^{K_2}_3| \nabla_{\bm{k}} H|\psi^{K_2}_1 \rangle & = 
-i\frac{\sqrt{3}}{\sqrt{2}}\omega^{-1},\\
 \bm{e}_{LCP}\cdot \langle\psi^{K_2}_3| \nabla_{\bm{k}} H|\psi^{K_2}_1 \rangle & = 0,\\
 \bm{e}_{RCP}\cdot \langle\psi^{K_2}_3| \nabla_{\bm{k}} H|\psi^{K_2}_2 \rangle & = 
\frac{\sqrt{3}}{\sqrt{2}}\omega^{-1},\\
 \bm{e}_{LCP}\cdot \langle\psi^{K_2}_3| \nabla_{\bm{k}} H|\psi^{K_2}_2 \rangle & = 0.
\end{align*}
Thus, only RCP light can excite electrons at $K_1$ points, i.e., {\it valley selective
optical excitation}. 
On the contrary, at $K_2$ points, only LCP light can excite electrons.
However, it is noted that the optical absorption is relatively weak
compared with those at $\Gamma$ and $M$ points, owing to the smaller JDOS
of Dirac cones.

\noindent
{\bf Absorption intensity}:
Figure~\ref{fig:figure6} (d) shows the energy dependence of absorption
intensity for the linearly and circularly polarized light irradiation,
together with the valley selective optical excitation. In upper pannel of
Fig.~\ref{fig:figure6} (d), the spectrum is integrated over the whole
1st BZ, and the sharp peaks at $\hbar\omega=2\gamma$ and $4\gamma$ 
are originated from optical transition at $M$ point, where JDOS diverges
logarithmically. 

It is intriguing that the lower peak at $2\gamma$ is dominated by the
optical transition $T_1$, as shown in middle panel of
Fig.~\ref{fig:figure6} (d). This behavior is attributed to the angle
dependence at $M$ point showed in Fig.~\ref{fig:figure6} (b).
Thus, the lower peak has strong momentum selectivity for the optical
absorption. For higher peak at $4\gamma$, no such strong selectivity is
found.  
Also, for circularly polarized light irradiation, such strong
 polarization dependence does not occur 
as shown in lower panel of Fig.~\ref{fig:figure6} (d). Instead,
the valley selective optical excitation is relatively enhanced around $\hbar\omega=3\gamma$.
Around this energy, optical transition occurs mainly at $K$ points between
Dirac cones and flat band. 
Although the optical absorption intensity is not so large
owing to the relatively small DOS near Dirac cones,
they show stronger valley selective optical exciation. 
\bibliographystyle{apsrev4-1} 
\bibliography{reference}
\end{document}